\documentstyle[12pt]{article}

    \let\p=\pi 
     
\let\w=\omega	   
  \let\S=\Sigma   
 
\let\la=\label  
  
 \def\bd{\begin{document}} \def\ed{\end{document}}
\def\ds{\documentstyle} \let\fr=\frac \let\bl=\bigl \let\br=\bigr
\let\Br=\Bigr \let\Bl=\Bigl
\let\bm=\bibitem
\let\na=\nabla
\let\pa=\partial \let\ov=\overline
\newcommand{\be}{\begin{equation}}
\newcommand{\ee}{\end{equation}}
\def\ba{\begin{array}}
\def\ea{\end{array}}
\def\S{{\bf S}}
\def\R{{{\bf R}}}
\def\Z{{\bf Z}}
\def\tilde{\widetilde}
\def\bar{\overline}
\newcommand{\ho}[1]{$\, ^{#1}$}
\newcommand{\hoch}[1]{$\, ^{#1}$}
\newcommand{\bea}{\begin{eqnarray}}
\newcommand{\eea}{\end{eqnarray}}
\newcommand{\ra}{\rightarrow}
\newcommand{\lra}{\longrightarrow}
\newcommand{\Lra}{\Leftrightarrow}
\newcommand{\ap}{\alpha^\prime}
\newcommand{\bp}{\tilde \beta^\prime}
\newcommand{\tr}{{\rm tr} }
\newcommand{\Tr}{{\rm Tr} }
\newcommand{\NP}{Nucl. Phys. }
\newcommand{\tamphys}{\it Center
for Theoretical Physics, Texas A\&M University, College Station, Texas 77843,
U. S. A.\\} 
\newcommand{\cern}{\it Theory Division, CERN, CH 1211 Geneva 23, Switzerland}
\newcommand{\princeton}{\it School of Natural Sciences, Institute for
Advanced Study, Olden Lane, Princeton, NJ 08540, U. S. A.} 
\newcommand{\auth}{M. J. Duff\footnote{
Research supported in part by NSF Grant PHY-9411543.}}
\newcommand{\another}{R. Minasian\footnote{World Laboratory Fellow.}}
\newcommand{\yetanother}{Edward Witten\footnote{Research supported
in part by NSF Grant PHY92-45317.}}
\begin{document}

\hfill{CTP-TAMU-54/95}

\hfill{hep-th/9601036}

\vspace{20pt}

\begin{center}
{ \large {\bf EVIDENCE FOR HETEROTIC/HETEROTIC DUALITY  }}

\vspace{24pt}

\auth

\vspace{10pt}

{\tamphys}

\vspace{10pt}

\another

\vspace{10pt}

{\cern}

\vspace{10pt}

and

\vspace{10pt}

\yetanother

\vspace{10pt}

{\princeton}

\vspace{24pt}

\underline{ABSTRACT}

\end{center}

We re-examine the question of heterotic - heterotic string duality
in six dimensions and argue that the $E_8\times E_8$ heterotic
string, compactified on $K3$ with equal instanton numbers in the two
$E_8$'s, has a self-duality that inverts the coupling, dualizes
the antisymmetric tensor, acts non-trivially on the hypermultiplets,
and exchanges gauge fields that can be seen in perturbation theory
with gauge fields of a non-perturbative origin.
The special role of the symmetric embedding of the anomaly
in the two $E_8$'s can be seen from field theory considerations
or from an eleven-dimensional point of view.
The   duality can be deduced by looking in two different ways
at eleven-dimensional $M$-theory 
compactified on $K3\times {\bf S}^1/\Z_2$.

\pagebreak
%\vskip10pt

\section{Introduction}
\la{Introduction}

Prior to the recent surge of interest in a duality between heterotic and
Type IIA strings  \cite{Hull,Witten,Senssd,Harvey,Rahmfeld3}, it was
conjectured (on the basis of $D=10$  heterotic string/fivebrane duality
\cite{Duffsuper,Strominger}) that in $D\leq6$ dimensions there ought to
exist a duality between one heterotic string and another
\cite{Luloop,Khurifour,Lublack,Minasian,Duffclassical,Khuristring,Duffstrong}. 
A comparison of the fundamental string solution \cite{Dabholkar} and the dual
solitonic string solution \cite{Lublack,Minasian} suggests the following $D=6$
duality dictionary: the dilaton $\tilde \Phi$, the string
$\sigma$-model metric $\tilde G_{MN}$ and $3$-form field strength $\tilde
H$ of the dual string are related to those of the fundamental string,  $\Phi$,
$G_{MN}$ and $H$ by the replacements      
\[
\Phi \rightarrow \tilde \Phi=-\Phi
\]
\[
G_{MN} \rightarrow \tilde G_{MN}=e^{-\Phi}G_{MN}
\] 
\be
H \rightarrow \tilde H=e^{-\Phi}*H
\la{dual}
\ee

In going from the fundamental string to the dual string, one
also interchanges the roles of worldsheet and spacetime loop expansions.
Moreover, since the dilaton enters the dual string equations with the opposite
sign to the fundamental string, it was argued in \cite{Luloop,Lublack,Minasian}
that in $D=6$ the strong coupling regime of the string should correspond to the
weak coupling regime of the dual string:  
\be
{\lambda}_6 = <e^{\Phi/2}>=1/{\tilde{\lambda}_6}
\la{coupling}
\ee
where $\lambda_6$ and $\tilde \lambda_6$ are the fundamental string and dual
string coupling constants. Because this duality
interchanges worldsheet and spacetime loop
expansions -- or because it acts by duality on $H$ -- the duality
exchanges the tree level Chern-Simons contributions to the Bianchi 
identity
\[
dH=\alpha'(2\pi)^2X_4 
\] 
\be
X_4=\frac{1}{4(2\pi)^2}[\tr R^2-\Sigma_\alpha v_\alpha \tr F_\alpha{}^2] 
\la{Bianchi} 
\ee
with the  one-loop Green-Schwarz corrections to the field equations
\[
d\tilde H=\alpha'(2\pi)^2\tilde X_4
\]
\be
\tilde X_4=\frac{1}{4(2\pi)^2}[\tr R^2-\Sigma_\alpha {\tilde v}_\alpha 
\tr F_\alpha{}^2]
\la{field}
\ee
Here $F_\alpha$ is the field strength of the $\alpha^{th}$ component
of 
the gauge group, $\tr$ denotes the trace in the fundamental representation, and
$v_\alpha,\tilde v_\alpha$ are constants.  (As explained in Appendix A, we may,
without loss of generality, choose the string tension measured in the string
metric and the dual string tension mesured in the dual string metric to be
equal.) 
In fact, the Green-Schwarz anomaly cancellation mechanism in six
dimensions requires that the anomaly eight-form $I_8$ factorize as a product of
four-forms, 
\be I_8=X_4\tilde X_4,
\la{eightform} 
\ee
and a six-dimensional string-string duality with the general features
summarized above would exchange the two factors.

Until now, there has not been a really convincing example of
heterotic-heterotic duality in six dimensions.  In  
\cite{Minasian}, it was proposed that the $D=10$ $SO(32)$ heterotic string
compactified to $D=6$ on $K3$ might be dual to the $D=10$ $SO(32)$ heterotic
fivebrane wrapped around $K3$. However, this candidate for a 
heterotic/heterotic
dual string pair suffered from the following drawbacks:

1) The existence of a fivebrane carrying the requisite $SO(32)$ quantum numbers
is still unclear.  Even if it exists, its properties are not well-understood.

2) The anomaly eight-form of this model is given by (\ref{eightform}) with
\cite{Erler}
\[
X_4=\frac{1}{4(2\pi)^2}[\tr R^2-\tr F_{SO(28)}{}^2-2\tr F_{SU(2)}{}^2] 
\]
\be
\tilde X_4 =   \frac{1}{4(2\pi)^2}[\tr R^2+2\tr F_{SO(28)}{}^2-44\tr
F_{SU(2)}{}^2], \la{SO(32)} 
\ee
and one of the gauge coefficients in the second factor enters with the wrong
sign.

The structure of this equation actually presents a problem that is
independent of any speculation about string-string duality.   
It was shown  by Sagnotti \cite{Sagnotti} that
 corrections to the Bianchi identities of the type (\ref{Bianchi}) and
to the field equations of the type (\ref{field}) are entirely consistent
with supersymmetry, with no restrictions on the constants
$v_\alpha$ and $\tilde v_\alpha$.  Moreover, supersymmetry
relates these coefficients to the gauge field kinetic energy.  In the Einstein 
metric $G^c{}_{MN}=e^{-\Phi/2}G{}_{MN}$, the
exact dilaton dependence of the kinetic energy of the gauge field
$F_\alpha{}_{MN}$, is 
\be 
L_{gauge}=-\frac{(2\pi)^3}{8\alpha'}\sqrt{G^c}\Sigma_\alpha \left( v_\alpha
e^{-\Phi/2}  +\tilde v_\alpha e^{\Phi/2} \right)\tr
F_\alpha{}_{MN}F_\alpha{}^{MN}. 
\la{happiness}
\ee
Positivity of the kinetic energy for all values of $\Phi$ thus implies that
$v_\alpha$ and $\tilde v_\alpha$ should both be non-negative, and at least one
should be positive. This fails for the $SO(32)$ heterotic string, as we see 
from
the formula for the anomaly eight-form.  Some interesting
new ``phase transition'' must occur at the value of $\Phi$ at which
the $SO(28)$ coupling constant appears to change sign, and at least until this
phase transition is understood, its occurrence might well
obstruct simple attempts to 
extrapolate from a string description at large negative $\Phi$ to a dual
string description at large positive $\Phi$.

In this paper, we shall attempt to remedy these problems as follows.

$1'$) It has recently been recognised that the ten-dimensional $E_8 \times E_8$
heterotic string is related to eleven-dimensional $M$-theory on ${\bf
R}^{10} \times \S^1/\Z_2$ \cite{Horava}, just as the ten-dimensional Type
IIA string is related to $M$-theory on ${\bf R}^{10}  \times \S^1$
\cite{Howe}.  By looking in two different ways at $M$-theory
on $\R^6\times K3\times \S^1/\Z_2$, we get a definite
framework for deducing string-string duality.  This framework shows
that the gauge group should be $E_8\times E_8$, the vacuum gauge bundle
should have equal instanton numbers in each $E_8$ (a situation we will
refer to as symmetric embedding
\footnote{Symmetric embedding
entered naturally 
 in \cite{Kachru} in constructing simple examples of 
heterotic/Type II duality in four dimensions.}), and the duality
acts in a non-trivial fashion on the hypermultiplets.

\def\T{{\bf T}}
{}From this eleven-dimensional point of view, one heterotic string comes by 
wrapping the $D=11$ membrane around $\S^1/\Z_2$ and the dual
heterotic string is obtained by reducing the $D=11$ fivebrane on
$\S^1/\Z_2$ and then wrapping around $K3$. 
This is quite similar to the eleven-dimensional derivation of
heterotic - Type IIA duality, which is recovered
if we replace $\S^1/\Z_2$ by $\S^1$ in the above scenario  
\cite{Liu}.\footnote{
The interpretation of the heterotic string as a  wrapping of a fivebrane
around $K3$, or around a $K3$ sub-manifold of a Calabi-Yau manifold or Joyce
manifold, is presumably the explanation for the ubiquity in string/string
duality of $K3$ itself and Calabi-Yau and Joyce manifolds corresponding to
fibrations of $K3$.}

$2'$) Now let us discuss the anomaly polynomial.  Picking a vacuum on $K3$ with
equal instanton numbers in  each $E_8$ will break $E_8\times E_8$
to a subgroup.  Generically $E_8\times E_8$ is completely broken,
so there are no questions of whether the gauge
contributions to the anomaly eight-form are compatible with
duality.  But we also want to understand how the duality acts
on vacua with non-trivial unbroken gauge groups.  For instance,
in a vacuum in which the gauge bundle breaks $E_8\times E_8$
to  $E_7\times E_7$
 (a maximal possible unbroken subgroup of $E_8\times E_8$) 
the anomaly eight-form is 
\be
I_8=\frac{1}{4(2\pi)^2}[\tr R^2-\frac{1}{6}\tr F_{E_7}^2-\frac{1}{6}\tr 
F_{E_7}^2]
\frac{1}{4(2\pi)^2}[\tr R^2] 
\la{symmetric}
\ee
We see that $\tilde v_\alpha=0$, so (i) there is no wrong sign problem
and one can possibly 
extrapolate to strong coupling without meeting a phase transition,
but (ii) since $v_\alpha\not=\tilde v_\alpha$, there is no manifest
self-duality.  Qualitatively similar results hold (that is, $\tilde 
v_\alpha=0$)
for any other unbroken subgroup of $E_8\times E_8$.  This qualitative
picture depends on having equal instanton numbers in the two $E_8$'s; 
in any other case, $\tilde v_\alpha<0$ for some subgroups
of $E_8\times E_8$, and phase transitions of some kind are unavoidable.

Because of (ii), it might appear that duality is impossible, but since
as in $1'$) above there is a systematic
framework for deducing the duality, we are    reluctant to accept this
interpretation.  We are led therefore to  assume that the duality
exchanges perturbative gauge fields, that is gauge fields of perturbative
origin, with non-perturbative gauge fields.  Despite the name,
non-perturbative gauge fields, if they appear at all, appear no matter how
small the string coupling constant may be; in fact, as the dilaton
is part of a tensor multiplet in $K3$ compactification, 
the unbroken gauge group is independent
of the string coupling constant at least if the low energy world can
be described by known physics.  Non-perturbative gauge fields, that is
gauge fields that are not seen in perturbation theory but appear
no matter how weak the string coupling constant may be,
can therefore only appear at points in moduli space at which perturbation
theory breaks down because of a kind of singularity.

A prototype for
non-perturbative gauge fields in the heterotic string
are the $SU(2)$ gauge fields that arise
for the $SO(32)$ heterotic string when an instanton shrinks to
zero size \cite{Wittensmall}.  Such gauge fields have $v_\alpha =0$;
this can be seen either from (a) the form of the anomaly polynomial,
as computed in section (4) of \cite{Schwarz}; (b) the physical picture
of \cite{Wittensmall} according to which making the heterotic string
dilaton smaller causes the $SU(2)$ gauge multiplet to appear ``farther
down the tube,'' without changing its physical properties such as the
gauge coupling; or (c) the description in \cite{Wittensmall} in terms
of  Type I $D$-branes,  where one can explicitly compute the $SU(2)$
gauge coupling and compare to (\ref{happiness}).\footnote{The vact
that $v=0$ for these non-perturbative gauge fields was in essence
also noted by V. Kaplunovsky.}  
In contrast to non-perturbative gauge fields which have $v_\alpha=0$,
 perturbative
gauge fields always have $v_\alpha>0$.  In fact, as discussed in Appendix
B, $v_\alpha$ is essentially the Kac-Moody level. 

Thus, our proposal is that heterotic string-string duality, at least
in the case of the symmetric embedding in $E_8\times E_8$, exchanges
perturbative gauge fields of $\tilde v_\alpha=0$
with non-perturbative  gauge fields of $v_\alpha=0$.
An interesting conspiracy of factors makes this perhaps
radical-sounding proposal possible.  (A) Since non-perturbative
gauge fields can only arise at particular loci  in hypermultiplet
moduli space (where a singularity develops in the $K3$ manifold or its
gauge bundle, giving a possible breakdown of perturbation theory
as in \cite{Wittensmall}), 
the proposal is possible only because with the symmetric
embedding in the gauge group, $E_8\times E_8$ is generically completely
broken, and  perturbative gauge fields only appear at particular loci
in hypermultiplet moduli space.  (B) Since the loci in hypermultiplet
moduli space which are candidates for nonperturbative gauge fields
(because of a singularity in the manifold or the gauge bundle)
are different from the loci where symmetry breaking is partly turned off
and unbroken perturbative gauge fields appear,
the proposal is possible only because the mechanism for string-string
duality alluded to in $1'$) gives a duality that acts non-trivially
on the hypermultiplets.
(C) It is essential that  with the
symmetric embedding promised in $1'$), one has $\tilde v_\alpha=0$ for
all perturbative gauge fields. 
If indeed  one had $\tilde v_\alpha<0$ in some case, one would have
to face the issue of the phase transition implied by the wrong sign
gauge kinetic energy.  On the other hand, perturbative gauge fields
with $\tilde v_\alpha>0$ would have to be dual to perturbative gauge fields,
leading to a contradiction given that one does not have manifest
duality of the perturbative gauge fields.

Once one accepts that after exchanging perturbative and non-perturbative
gauge fields, the $v_\alpha$ and the
$\tilde v_\alpha$ are equal,  the equality of the numerical coefficents
appearing in the gauge kinetic terms (\ref{happiness}), and in the field
equations and Bianchi identities (\ref{field}), (\ref{Bianchi}) means 
that there may now be a full-fledged 
{\it self-duality} of the $D=6$ string extending the symmetry of the
low energy supergravity and acting on some of the massless fields by
(\ref{dual}):
\[ 
\Phi \rightarrow -\Phi
\]
\[
G_{MN} \rightarrow e^{-\Phi}G_{MN} 
\] 
\be
H \rightarrow e^{-\Phi}*H
\la{discrete}
\ee

 \section{The	Fundamental	String on $K3$ And
The Low Energy Supergravity} \label{K3}

$K3$ compactification of a chiral string theory in ten dimensions
gives a chiral six-dimensional theory.
$K3$	is	a four-dimensional	compact	closed
simply-connected	manifold.	It	is equipped	with	a 
self-dual	metric	with	holonomy group		$SU(2)$.
It	was	first	considered 	in	a
Kaluza-Klein	context	in \cite{Pope,Townsend} where	it	was	used,
in particular,	as	a	way	of	compactifying	$D=11$	
supergravity to	$D=7$ and
$D=10$	supergravity to	$D=6$.	
Our interest here is in $K3$ compactification of the heterotic string.
Because 	of the	$SU(2)$	holonomy,	
half the supersymmetry survives in $K3$ compactification,
and
	hence, starting from
$N=1$ supergravity in $D=10$, we get  $N=1$ supergravity in $D=6$
(which has half as many supercharges).

There	are	four	massless 
$N=1,D=6$	supermultiplets	to	consider: 
$$
\begin{array}{ll}
Supergravity~Multiplet
~~~~~&G_{MN},	\Psi^{A+}_{M},	B^+{}_{MN}\\
Tensor~Multiplet~~~~~~~~~~~~&B^-{}_{MN},	\chi^{A-},	\Phi\\
Hypermultiplet~~~~~~~&\psi^{a-},	\phi^{\alpha}\\
Yang-Mills~Multiplet~~~~~~~~&A_{M},	\lambda^{A+}
\end{array}
	$$
All	spinors	are
symplectic	Majorana--Weyl.	The	
two-forms	$	B^+{}_{MN}$	and
$B^-{}_{MN}$
have	three-form	
field	strengths	that	are	self-dual	and
anti-self-dual,
respectively.	Only with precisely one tensor multiplet
added to the supergravity multiplet	is there	a	conventional	
covariant
Lagrangian	formulation.
In $K3$ compactification,	the	zero modes	of the
supergravity	multiplet	in	$D=10$	give 
this	combination	plus	$20$ massless
matter	hypermultiplets.	The	$80$	scalars	in those
multiplets parametrize the 	coset	$SO(20,4)/{SO(20)
\times	SO(4)}$
\cite{Nilsson1,Seiberg,Aspinwall1}, which is 	the
moduli	space	of	conformal field theories on $K3$.
No 	vector	multiplets come from the ten-dimensional supergravity
multiplet 	since	$K3$	has	no
isometries	and	is	simply connected. 

Six-dimensional vector multiplets and additional hypermultiplets
come from reduction on $K3$ of the ten-dimensional gauge group
$SO(32)$ or $E_8\times E_8$, as was first analyzed in \cite{Green2}.
An important constraint comes from the anomaly cancellation
equation $dH=(\alpha'/4)\left(\tr R^2-\sum_\alpha v_\alpha \tr F_\alpha^2
\right)$ which was discussed in the Introduction.  A global solution
for $H$ exists if and only if the integral over $K3$ of $\tr R^2$
equals that of $\sum_\alpha v_\alpha \tr F_\alpha^2$.  This
amounts to the statement that the vacuum expectation value of the $SO(32)$
or $E_8\times E_8$ gauge fields must be a configuration with instanton
number 24.  In the $E_8\times E_8$ case, it is the sum of the instanton
numbers in the two $E_8$'s that must equal 24.  We will be interested
mainly in the ``symmetric embedding,'' the case where the instanton
number is 12 in each $E_8$.

With instanton number 12 in each $E_8$, the generic $E_8\times E_8$
instanton on $K3$ completely breaks the gauge symmetry.  This will
be explained in more detail in section (4), together with generalizations.
Unbroken gauge symmetry arises if the vacuum gauge bundle takes values
in a subgroup  $G$ of $E_8\times E_8$, in which case the unbroken
subgroup of $E_8\times E_8$ is the commutant $G$ of $H$, that is the
subgroup of $E_8\times E_8$ that commutes with $H$.
The $G$ quantum numbers of the massless hypermultiplets can be determined
by decomposing the adjoint representation of $E_8\times E_8$ under
$G\times H$, as in \cite{Green2}.  

\bigskip
\noindent{{\it The Anomaly Polynomial}}

Now let us explain some statements made in the Introduction about the
anomaly polynomials.  Let $F_i$, $i=1,2$ be the field strengths of
the two $E_8$'s.  Let $\Tr F_i{}^2$ be the traces in the adjoint
representations of the two $E_8$'s, and let $\tr F_i{}^2=(1/30) \,\Tr F_i{}^2$.
In ten dimensions, the anomaly twelve-form $I_{12}$ factorizes
as $I_{12}=X_4\tilde X_8$, with
\be
X_4=\frac{1}{4(2\pi)^2}[\tr R^2-\sum_i\tr F_i{}^2]
\ee
and $\tilde X_8$ the more imposing expression  
\[
\tilde X_8 \sim {3\over 4}\left((\tr F_1{}^2)^2+(\tr
F_2{}^2)^2\right) -{1\over 4}\left(\tr F_1{}^2+\tr F_2{}^2\right)^2
\]
\be
-{1\over 8}\tr R^2
\left(\tr F_1{}^2+\tr F_2{}^2\right)
+{1\over 8}\tr R^4+{1\over 32}
\left(\tr R^2\right)^2.
\ee

The six-dimensional anomaly four-form $\tilde X_4$ is obtained by
integrating $\tilde X_8$ over $K3$.  
Also in writing an anomaly four-form in six dimensions, one understands
the six-dimensional field strengths
$F_1$ and $F_2$ to take values in
the unbroken subgroup of the gauge group, that is
the part that commutes with the gauge bundle on $K3$.
If we let $\langle R^2\rangle$,
$\langle F_1{}^2\rangle$, and $\langle F_2{}^2\rangle$ denote the integrals
of $\tr R^2$, $\tr F_1{}^2$, and $\tr F_2{}^2$ over $K3$, then 
\[
\tilde X_4 \sim \tr F_1{}^2\left({1\over 2}\langle F_1{}^2\rangle
-{1\over 4}\langle F_2{}^2\rangle -{1\over 8}\langle R^2\rangle\right)
\]
\be
+\tr F_2{}^2\left({1\over 2}\langle F_2{}^2\rangle-{1\over 4}\langle
F_1{}^2\rangle -{1\over 8}\langle R^2\rangle \right)
+\tr R^2
\left({1\over 16}\langle R^2\rangle -{1\over 8} (\langle F_1{}^2\rangle
+\langle F_2{}^2\rangle)\right).
\ee
The topological condition on the vacuum gauge bundle that was
explained above amounts to
\be 
\langle F_1{}^2\rangle +\langle F_2{}^2\rangle -\langle R^2\rangle =0.
\ee
With the use of this equation, one sees that the coefficients $\tilde v_i$ 
of $\tr F_1{}^2$ and $\tr F_2{}^2$ in $\tilde X_4$ are equal and opposite.
So the ``wrong sign'' problem explained in the Introduction is avoided
if and only if the $\tilde v_i$ vanish.  (Otherwise, the  problem
arises in the $E_8$ that has smaller instanton number.)
{}From the above formulas, 
the condition for this is that $\langle F_1{}^2\rangle
=\langle F_2{}^2\rangle$, that is the two $E_8$ gauge bundles have
equal instanton numbers.  So we have recovered the statement in the 
Introduction that in $E_8\times E_8$, the sign problem is avoided only
for the symmetric embedding, for which the $\tilde v_i$ vanish for all
subgroups of $E_8\times E_8$.  A similar analysis for $SO(32)$ recovers
the anomaly formula given in the Introduction and in particular
shows the occurrence of the sign problem for $SO(32)$.

\bigskip
\noindent{{\it Further Aspects Of The Low Energy Supergravity}}

As a guide to the kind of dualities one might expect in the string
theory, let us look in more detail at the corresponding $N=1,D=6$
supergravity theories.  We shall follow \cite{Sagnotti} but with the
following modifications: by choosing just one tensor multiplet we may write a
covariant Lagrangian as well as covariant field equations; we use the string
$\sigma$-model metric so as to emphasize the tree-level plus one-loop nature
of the Lagrangian; we also write the coupling in terms of the string slope
parameter $\alpha'$; we shall also include {\it Lorentz} as well as Yang-Mills
corrections to the Bianchi indentities and field equations.  We shall denote 
the
contribution to the action of $L$ fundamental string loops and $\tilde L$ dual
string loops by $I_{L,\tilde L}$. The bosonic part of the action of the takes
the form            
\be   
I=I_{00}+I_{01}+I_{10} +...  
\ee 
 where
\[
I_{00}=\frac{(2\pi)^3}{\alpha'^2}\int d^6x \sqrt{-G}e^{-\Phi}[R_G+G^{MN} 
\partial_M\Phi\partial_N\Phi 
\]  
\be 
-\frac{1}{12}G^{MQ}G^{NR}G^{PS}H_{MNP}H_{QRS}]
\la{00}
\ee
where $M,N=0,...,5$ are spacetime indices and $H$ is the curl of a $2$-form
$B$, where  
\[
I_{01}=\frac{(2\pi)^3}{8\alpha'}\int d^6x
\sqrt{-G}e^{-\Phi}[G^{MP}G^{NQ}\tr R_{MN}R_{PQ}
\]
\be
-\Sigma_\alpha v_\alpha G^{MP}G^{NQ}\tr {F_\alpha}_{MN}{F_\alpha}_{PQ}]
\la{01} 
\ee
together with Chern-Simons corrections to $H$ appropriate to (\ref{Bianchi}),
and where 
\[ 
I_{10}=\frac{(2\pi)^3}{8\alpha'}\int d^6x
\sqrt{-G}[G^{MP}G^{NQ}\tr R_{MN}R_{PQ}
-\Sigma_\alpha \tilde v_\alpha G^{MP}G^{NQ}\tr {F_\alpha}_{MN}{F_\alpha}_{PQ}
\]
\be
-	2\p	\int_{M_{6}}	\Bl	(	\frac{1}{(2\pi)^2 \alpha'}	\,	B\,
	\tilde	X_4	+	\frac{1}{3}\,	\w_3	\,\tilde	\w_3	\Br	)
\la{10}
\ee
Here $\w_3$ and $\tilde	\w_3$ obey $2d\w_3=X_4$ and $2d\tilde	\w_3=\tilde
X_4$. This last term ensures that $H$ obeys the field equations appropriate
to (\ref{field}). The metric $G_{MN}$ is related to the canonical Einstein
metric $G^c{}_{MN}$ by 
\be
G_{MN}=e^{\Phi/2}G^c{}_{MN}
\la{stringmetric}
\ee
where $\Phi$ the $D=6$ dilaton.  There will also be couplings to the
hypermultiplets, both charged and neutral, which we shall not attempt to write
down.  They will belong to some quaternionic manifold, which is probably quite
complicated. 

The most obvious dual supergravity action is given by a similar
expression obtained by replacing each field with its dual counterpart according
to the following duality dictionary:    
\[ 
\tilde \Phi=-\Phi 
\] 
\[ 
\tilde G_{MN}=e^{-\Phi}G_{MN}
\]  
\[ 
\tilde H=e^{-\Phi}*H 
\] 
\be 
\tilde A_M= A_M
\la{dictionary}
\ee
where $*$ denotes the Hodge dual.  (Since the $H$ equation is conformally
invariant, it is not necessary to specify which metric is chosen in forming the
dual.)   The dual metric $\tilde G_{MN}$ is related to the canonical Einstein
metric by  
\be
\tilde G_{MN}=e^{-\Phi/2}G^c{}_{MN}
\la{dualstringmetric}
\ee
It is also possible (and will be necessary in our application) to combine
the duality just described with a transformation of the hypermultiplets
and a permutation of the various possible factors in the gauge group:
\be
\alpha \rightarrow \pi(\alpha)
\ee  
where $\pi(\alpha)$ is the gauge group into which the duality maps the gauge
group $\alpha$. With or without such a transformation of hypermultiplets, the
above dictionary achieves just the right interchange of tree-level and one loop
effects required by heterotic/heterotic duality, namely   
\be
I_{10} \leftrightarrow I_{01}
\ee
In particular, with $\tilde H$ the field strength of a two-form $\tilde B$, 
with
Chern-Simons corrections appropriate to (\ref{field}), this duality
exchanges the  Bianchi identities (\ref{Bianchi}) and field equations
(\ref{field}).
When the permutation $\alpha\to\pi(\alpha)$ is taken into account, we
have hopefully
\be
v_\alpha=\tilde v_{\pi(\alpha),}
\ee
reflecting the discrete symmetry (\ref{discrete}).

\section{Deduction Of String-String Duality From Eleven Dimensions}

To {\it deduce} heterotic-heterotic duality on $K3$, we begin
with the  eleven-dimensional $M$-theory on ${\bf R}^6\times K3 \times
\S^1/\Z_2$.  By looking at this theory in two different ways, we
will deduce a duality between heterotic strings.
On the one hand, we use the fact that the $M$-theory on $Y\times \S^1/\Z_2$,
for any $Y$, is equivalent to the $E_8\times E_8$ heterotic string
on $Y$, with a string coupling constant that is small as the  radius
of the $\S^1/\Z_2$ shrinks.  On the other hand, we use the fact that
the $M$-theory on $Z\times K3$, for any $Z$, is equivalent to the
heterotic string on $Z\times \T^3$, with a string coupling constant
that is small when the  $K3$ shrinks.
The point of starting with
$W={\bf R}^6\times K3\times \S^1/\Z_2$ is that it can be written
as either $Y\times \S^1/\Z_2$, with $Y=\R^6\times K3$, or as 
$Z\times K3$, with $Z=\R^6\times \S^1/\Z_2$. 

If we look at $W$ as $Y\times \S^1/\Z_2$, then we deduce that as
the $\S^1/\Z_2$ becomes small, the $M$-theory on $W$ is equivalent to the
$E_8\times E_8$ heterotic string on $Y=\R^6\times K3$. 

A little more subtlety is required if we try to look at $W$ as $Z\times K3$.
{}From this vantage point, it appears that as the $K3$ shrinks, we should
get a weakly coupled heterotic string on $Z\times \T^3=\R^6\times \S^1/\Z_2
\times \T^3$.  This cannot be the right answer, as $\R^6\times \S^1/\Z_2
\times \T^3$ is unorientable, and the parity-violating heterotic string
cannot be formulated on this space.  One must note that when
one divides $\R^6\times \S^1\times K3$ by $\Z_2$ to get 
the $M$-theory on $\R^6\times \S^1/\Z_2\times K3$, the three-form
potential $A$ of the low energy limit of the $M$-theory is odd under
the $\Z_2$.  Compactified on $K3$, the three-form gives 22 vector
fields that come from the two-dimensional cohomology of $K3$; these
are related to the momentum and winding modes and  Wilson lines of the
heterotic string on $\T^3$.  For all the momentum and winding modes
of the heterotic string to be odd under the $\Z_2$ means that 
the $\Z_2$ must act as $-1$ on the $\T^3$.  So when we shrink the
$K3$ factor in the $M$-theory  on $\R^6\times \S^1/\Z_2\times K3$,
we get a heterotic string on not $\R^6\times \S^1/\Z_2\times \T^3$ 
but $\R^6\times \left(\S^1\times \T^3\right)/\Z_2$.

Now roughly speaking $\left(\S^1\times \T^3\right)/\Z_2=\T^4/\Z_2$
is a $K3$ orbifold, so we have arrived again at a heterotic string
on $K3$.  Actually, there are a few subtleties hidden here.  For one
thing, in general we will not really get in this way a $K3$ orbifold.
When the $M$-theory is formulated on $\R^6\times \S^1/\Z_2\times K3$, there
are propagating $E_8$ gauge fields on
each copy of $\R^6\times K3$ coming from a fixed point in the $\Z_2$
action on $\S^1$,  and a specification of vacuum requires picking
a $K3$ instanton on each $E_8$.  A choice of such an instanton represents,
at least generically, a departure from a strict orbifold vacuum.
Since the vacuum on $\R^6\times \S^1/\Z_2\times K3$ was not really
an orbifold vacuum, the same will be true by the time we get to
a heterotic string on $\R^6\times \left(\S^1\times \T^3\right)/\Z_2$.

The fact that $M$-theory on $\R^6\times \S^1/\Z_2\times K3$ turns
into a weakly coupled heterotic string in  two different limits
is a kind of duality between heterotic strings.  From the point of
view of either one of these limits, the other one is strongly coupled; we
will be more precise about this below.  An observer
studying one of the two limiting heterotic strings sees a strongly
coupled limit in which  there is a weakly coupled description
by a different heterotic string;  this is heterotic-heterotic duality.

A few points should still be explained: {\it (i)} The duality is an
electric-magnetic string-string duality in the sense described in the
Introduction. {\it (ii)} The duality acts non-trivially on the hypermultiplets,
a fact whose importance was explained in the Introduction.
{\it (iii)}  The construction -- which obviously requires that the
heterotic string gauge group be $E_8\times E_8$ -- works only 
for the symmetric embedding
with equal instanton numbers in the two $E_8$'s.

The first point is a simple consequence of eleven-dimensional facts.
\footnote{The following two paragraphs benefited from a discussion with
P. Horava.}
We begin with the fact that the eleven-dimensional $M$-theory has
two-branes and five-branes that are electric-magnetic duals.
Consider in general a Kaluza-Klein vacuum in a 
theory containing a $p$-brane; 
for simplicity consider the illustrative case that the vacuum
is $Q\times \S^1$ for some $Q$.  A $p$-brane can be wrapped around
$\S^1$, giving a $(p-1)$-brane on $Q$. Or a $p$-brane can be ``reduced''
on $\S^1$, by which we mean simply that one takes the $p$-brane to
be localized at a point on $\S^1$.  This gives a $p$-brane on $Q$, with
the position on $\S^1$ seen an a massless world-volume mode.
The two operations of wrapping and reduction are electric-magnetic duals,
so that if one starts with dual $p$-branes and $q$-branes, the wrapping
of one around  $\S^1$ and reduction of the other on  $\S^1$ gives 
dual objects on $Q$.

Now let us apply this wisdom to $M$-theory on $\R^6\times \S^1/\Z_2\times K3$.
In eleven dimensions, the $M$-theory has dual two-branes and five-branes.
When the $\S^1/\Z_2$ shrinks, an effective heterotic string in ten dimensions
is obtained by wrapping the two-brane around $\S^1/\Z_2$, giving a
 one-brane which was seen in \cite{Horava} to have the world-sheet
structure of an $E_8\times E_8$  heterotic string.  An effective
six-dimensional heterotic string is then obtained by reduction on $K3$.
On the other hand, when the $K3$ shrinks, an effective heterotic string
is obtained by wrapping the five-brane around $K3$.  An effective
six-dimensional heterotic string is  then obtained by reduction on $\S^1/\Z_2$.
Since wrapping the five-brane around $K3$ and reducing it on $\S^1/\Z_2$ is 
dual
to wrapping the membrane around $\S^1/\Z_2$ and reducing it on $K3$, the
two effective six-dimensional heterotic strings 
are electric-magnetic duals to each other in the sense described
in the Introduction.  This provides an answer to question {\it (i)} above.
To further confirm our understanding,
we compute below the six-dimensional string coupling
constants of the two heterotic string theories, and show that they are inverses
of each other, as expected for a pair of six-dimensional dual strings.

Now we come to question {\it (ii)}, which is to show that this duality
is not the minimal duality suggested  by low energy supergravity, but
acts non-trivially on the hypermultiplets.  In fact, begin with
$M$-theory on $\R^6\times \S^1/\Z_2\times K3$, with some modulus for the 
$K3$ and with a  particular choice of $E_8$ gauge bundles at fixed points.  
If one shrinks
the $\S^1/\Z_2$ one simply gets a heterotic string on the {\it same}
$K3$, with the same $E_8$ gauge bundles, that one started with.
If instead one shrinks the $K3$, one has an adventure described above
involving a non-orbifold vacuum on $\left(\S^1\times \T^3\right)/\Z_2$.
This is presumably to be interpreted as a $K3$ (in fact, any $(0,4)$
conformal field theory of the appropriate central charge is believed
to describe a $K3$ with a vector bundle), but it certainly  does not
look like the $K3$ that we started with.\footnote{
We have not yet given any explicit argument that the dual heterotic
string is an $E_8\times E_8$ theory with symmetric embedding.
That the gauge group of the dual string is
 $E_8\times E_8$ rather than $SO(32)$ we infer
from the fact that, if one starts with suitable gauge bundles in eleven
dimensions, unbroken exceptional gauge
groups such as $E_7\times E_7$ are possible.
Moreover, if one starts with the symmetric embedding in eleven dimensions
-- whose necessity we argue for presently -- then there is a symmetry
of $\S^1/\Z_2$ that exchanges the two fixed points and the two $E_8$'s.
This will carry over in the dual heterotic
string theory to a symmetry that exchanges
the two $E_8$'s, and the existence of this symmetry indicates
that the dual heterotic string has equal instanton numbers in the
two $E_8$'s.}
  We take this to mean that the
$K3$ associated with the dual string (obtained by wrapping the five-brane)
is {\it not} the same as the $K3$ we started with in eleven-dimensions,
or differently put that the duality acts non-trivially on the hypermultiplets,
which are the moduli of $K3$ and the vector bundle.  In fact,
the action on the hypermultiplets looks rather complicated, and understanding
it better would be an important step.

Finally we come to question {(\it iii)}.  As we have presented the
eleven-dimensional construction so far, the assignment of instanton
numbers to the two $E_8$ gauge bundles does not seem to matter.
But we claim that actually, if examined more closely, the construction
works only for the symmetric embedding, that is for equal instanton numbers
in the two $E_8$'s.  The three-form potential of eleven-dimensional
supergravity has a four-form field strength $K$.  $M$-theory compactifications
on $K3$ can be distinguished according to the quantized value of the flux
\cite{Liu}
\be
\int_{K3} K=\frac{2 \pi m}{T_3},~~~~~~m=integer
\ee  
where $T_3$ is the membrane tension.  $K3$ compactification of $M$-theory has
usually been discussed only for $m=0$.  In particular, the statement that when
a $K3$ shrinks, one gets a heterotic string with the $K3$ replaced by $\T^3$
holds for $K3$'s with $m=0$.  Intuitively, one would expect $m\not=0$ to change
the behavior that occurs
when one tries to shrink a $K3$, because the energy stored
in the trapped $K$ field would resist this shrinking.  

One can actually be more precise.  The dual heterotic string that
arises when one shrinks a $K3$ comes from a five-brane wrapped around
the $K3$.  But a five-brane cannot wrap around a $K3$ (or any four-manifold)
that has $m\not= 0$.  The reason for this is that the world-volume 
spectrum of the five-brane includes a massless two-form with
an anti-self-dual three-form field strength $T$.  $T$ does not obey
$dT=0$, but rather (as one can see from equation (3.3) of
\cite{Newtownsend};  another argument is given in \cite{Newwitten}) 
it obeys
$dT=K$.  The existence of a solution for $T$ means that $K$ must be
cohomologically trivial when restricted to the five-brane world-volume;
that is, the five-brane cannot wrap around a four-manifold with $m\not= 0$.

The reason that this is relevant is that, as we will argue momentarily,
if one compactifies the $M$-theory on $\R^6\times \S^1/\Z_2\times K3$ with
instanton number $k$ in one $E_8$ and $ 24-k$ in the other $E_8$,
then the flux of the $K$ field over $K3$ (that is, over any $K3$ obtained
by restricting to a generic point in $\R^6\times \S^1/\Z_2$) is
$m=\pm (12-k)$.  (The sign will be explained later.)  
Therefore, $m=0$ if
and only if $k=12$, that is, precisely for the symmetric embedding.
 The eleven-dimensional explanation of heterotic - heterotic
duality thus requires $E_8\times E_8$ with the symmetric embedding.

It remains, then, to explain the relation $m=\pm (12-k)$.  This relation
arises upon  writing the anomaly cancellation condition
\be
dH={\alpha'\over 4}\left(\tr R\wedge R -\tr F_1\wedge F_1-\tr F_2\wedge F_2
\right)
\la{anomeq}
\ee
 in 
eleven-dimensional terms.   The eleven-dimensional version
of that equation must involve the five-form $dK$ instead of the four-form
$dH$.  This requires incorporating on the right hand side
delta-functions supported at the fixed points.  
If $\S^1$ is parametrized by an angular variable 
$x^{11}$ such that the $\Z_2$ fixed points are at $x^{11}=0$ and $\pi$,
with $F_1$ supported at the first and $F_2$ supported at the second,
then the eleven-dimensional version of (\ref{anomeq}) is
\[
dK=\frac{1}{2\pi T_3}dx^{11} \times
\]
\be
 \left(\delta(x^{11})\left({1\over 2}\tr R\wedge
R-\tr F_1\wedge F_1\right) 
+ \delta(x^{11}-\pi)\left({1\over 2} \tr R\wedge R 
-\tr F_2\wedge F_2\right)\right).
\la{highanomeq}
\ee
This equation is determined by the following properties:
$dK$ vanishes except at fixed points, since (in the absence of five-branes)
$dK=0$ in the eleven-dimensional theory; $F_1$ and $F_2$ contribute
only at the appropriate values of $x^{11}$; the two fixed points
enter symmetrically; if one integrates over 
$x^{11}$ and interprets $H$ as the part of the zero mode of $K$ with
one index equal to 11, then (\ref{highanomeq}) reduces to (\ref{anomeq}).
Now, let $m(x^{11})$ be the function obtained by integrating
$T_3F/2\pi$ over $K3$ at a given value of $x^{11}$.  The $\Z_2$ symmetry
implies that $m(-x^{11})=-m(x^{11})$, and (\ref{highanomeq})
means that $m(x^{11})$ is constant 
except for jumps at $x^{11}=0$ or $\pi$,
the magnitude of the jump being $(2/8\pi^2)\int_{K3}\left({1\over 2}
\tr R\wedge R - \tr F_1\wedge F_1\right) 
=-(2/8\pi^2)\int_{K3} \left({1\over 2}
\tr R\wedge R -\tr F_2\wedge F_2\right)$.  Hence the constant value of
$m(x^{11})$ away from a fixed point, which we earlier called $\pm m$,
is 
\[
\pm  (1/8\pi^2) \int_{K3} \left({1\over 2}\tr R\wedge R
-\tr F_1\wedge F_1\right).
\]
 This amounts to the statement that
$m=\pm (12-k)$, with $k$ the instanton number in the first
$E_8$, supported at $x^{11}=0$.  This confirms the claim made
above and so completes our explanation of why
the eleven-dimensional approach to heterotic - heterotic duality
requires the symmetric embedding as well as requiring gauge group
$E_8\times E_8$. 

\bigskip\noindent{{\it Anomaly Cancellation By Five-Branes}}

We cannot resist mentioning an application of these ideas that is somewhat
outside our main theme.  The equation (\ref{highanomeq}) shows that
the curvature $\tr \,R\wedge R$ of $K3$ gives a magnetic source for
the $K$ field, that is a contribution to $dK$ supported at fixed points.
This is a sort of ``anomaly'' that must be canceled, since the integral
of $dK$ over the compact space $K3\times \S^1/\Z_2$ will inevitably
vanish.  The conventional string theory way to cancel this anomaly
is to use $E_8\times E_8$ instantons on $K3$, using the fact that
$-\tr \,F_i\wedge F_i$ also contributes to $dK$.
{}From this point of view, the ``magnetic charge'' associated with
the $\tr\,R\wedge R$ contribution to $dK$ can be canceled by 24 
$E_8\times E_8$ instantons.

There is, however, another standard entity that can contribute 
to $dK$; this is the
eleven-dimensional five-brane, which is characterized by the fact that 
$dK$ has a quantized delta-function contribution supported on the five-brane
world-volume.  This suggests that instead of canceling the total contribution
to $dK$ with 24 $E_8\times E_8$ instantons, one could use 23 instantons
on the fixed points and one five-brane at some generic point in 
$K3\times \S^1/\Z_2$.  More generally, one could use $24-n$ instantons
(distributed as one wishes between the two $E_8$'s) and $n$ five-branes.

The following is a strong indication that this is correct.
Supported on the five-brane world-volume is one tensor multiplet and
one hypermultiplet in the sense of $N=1$ supergravity in $D=6$.
(The hypermultiplet parametrizes the position of the five-brane on $K3$.)
Instead, associated with each $E_8\times E_8$ instanton are precisely
30 hypermultiplets (a number that can be seen in the instanton dimension
formula with which we begin the next section).  Both the tensor multiplet
and the hypermultiplet contribute to the irreducible part of the gravitational
anomaly in six dimensions (the part that cannot be canceled by a
Green-Schwarz mechanism).  From equation (118) of
\cite{Alv}, one can see that the contribution to this irreducible anomaly
of one tensor multiplet and one hypermultiplet equals that of 30 
hypermultiplets, strongly suggesting that the $M$-theory vacua
with $24-n$ instantons and $n$-fivebranes (and therefore $n+1$ tensor
multiplets in six dimensions) really do exist.  These cannot
be related to perturbative heterotic strings, but they might have
limits as Type I orientifolds, as in \cite{Sagnotti}.
They are somewhat reminiscent of the $M$-theory vacua with wandering
five-branes found in \cite{Newwitten}.  

\bigskip\noindent{{\it Analysis Of The Couplings}

We now return to the $M$-theory on $\R^6\times \S^1/\Z_2\times K3$,
with the intention of examining somewhat more quantitatively
the two limits in which it is related to a heterotic string  -- the
limits in which the $\S^1/\Z_2$ or the $K3$ shrinks to small volume.
We want to show that in either such limit, the heterotic string that
emerges is {\it weakly coupled}. In fact, part of the meaning of any claim
that the $M$-theory turns into a string theory in a particular limit
should be that the resulting string theory is weakly coupled.   
We will also, more precisely, show that the coupling constant
of the heterotic string obtained by shrinking the $\S^1/\Z_2$ is the
inverse of the coupling of the
 heterotic string obtained by shrinking the $K3$.
This is what one would expect given that these strings are electric - magnetic
duals.  The calculations we will need are quite straightforward given
formulas in \cite{Witten}.  In these computations we will not keep
track of some absolute constants.

We let $R$ be the radius of $\S^1/\Z_2$, measured with respect to
the metric of eleven-dimensional supergravity, and we let $V$ be the volume
of the $K3$, likewise measured in eleven-dimensional terms.  According to
\cite{Horava,Witten}, as $R$ goes to zero with large $V$,
one gets a heterotic string with the ten-dimensional string coupling
constant being $\lambda_{10} = R^{3/2}$.  Also, the string metric
differs from the 
eleven-dimensional metric by a Weyl rescaling, such that the $K3$ volume
measured in the string metric is $V_{st}=VR^2$.  The six-dimensional
string coupling constant $\lambda_6$ obeys the standard relation
$1/\lambda_6^2=V_{st}/\lambda_{10}
^2$, 
so we get
\be
\lambda_6^2={\lambda_{10}^2\over V_{st}} = {R\over V}.
\ee
This shows that the six-dimensionl string coupling constant is
small if $R<<V$.

Now we consider the opposite limit in which the $K3$ shrinks.
Then one gets in seven dimensions a dual heterotic string $\R^6\times
\S^1/\Z_2$ 
with (according to formulas in \cite{Witten}) a seven-dimensional
string coupling constant $\tilde \lambda_7 =V^{3/4}$.  Also,
the string metric of the dual heterotic string theory differs
by a Weyl rescaling from the eleven-dimensional metric, such that
the radius of the $\S^1/\Z_2$ in the dual string metric is
$\tilde R_{st} = V^{1/2}R$.  The six-dimensional dual string
coupling constant obeys
\be
\tilde \lambda_6^2={\tilde \lambda_7^2\over R_{st}}={V\over R}.
\ee
Putting these formulas together, we get $\lambda_6=1/\tilde\lambda_6$,
as expected from the fact that the string obtained by wrapping the
two-brane around $\S^1/\Z_2$ is dual to the string obtained by wrapping
the five-brane around $K3$.

\bigskip\noindent{{\it Behavior Under Further Compactification}}

It is interesting to consider further toroidal compactification to four
dimensions, replacing $\R^6$ by $\R^4\times \T^2$.  Starting with
a $K3$ vacuum in which the $E_8\times E_8$ gauge symmetry is completely
Higgsed, the toroidal compactification to four dimensions gives an
$N=2$ theory with the usual three vector multiplets $S$, $T$ and $U$ related
to the four-dimensional heterotic string coupling constant and
the area and shape of the $\T^2$.  When reduced to four dimensions,
the six-dimensional string-string duality (\ref{discrete}) becomes
\cite{Duffstrong} an operation that exchanges $S$ and $T$.  Since the heterotic
string on $\T^2\times K3$ also has $R\to 1/R$ symmetries that exchange
$T$ and $U$, this is an example with complete $S-T-U$ triality 
symmetry, as discussed in \cite{Rahmfeld3}.

Kachru and Vafa \cite{Kachru} made a proposal for a Type II dual
of the $E_8\times E_8$ heterotic string on $\T^2\times K3$ 
with this precise vacuum, that is equal instanton numbers in the
two $E_8$'s and complete Higgsing.  Some evidence for the $S-T$
interchange symmetry has appeared in subsequent study of this example
\cite{Klemm,Duffelectric,Cardoso}.

\section{The Duals Of Some Unbroken Gauge Groups}

The
moduli space ${\cal M}_k(E_8)$ of $E_8$ instantons on $K3$ with instanton 
number
$k$ has a dimension (predicted from the index formula) which is
\be
\dim\,{\cal M}_k(E_8) = 120 k - 992
\la{dimform}
\ee
if $k$ is sufficiently big.  
We will make frequent use of the special case
\be
\dim\,{\cal M}_{12}(E_8)=448.
\la{ugform}
\ee
The formula for $\dim\,{\cal M}_k$ actually
gives the  the correct
dimension  of the moduli space
if $k$ is large enough that a $K3$ instanton of instanton number
$k$ can completely break the $E_8$ gauge symmetry.  A necessary
condition for this to be possible is that the right hand side
of (\ref{dimform}) must be positive, restricting us to $k\geq 9$.  
We have checked that complete Higgsing is possible for $k \geq 10$ and
do not know if it is possible for $k=9$.\footnote{The check for $k=10$
can be made for instance by starting with instanton number 10 in
an $SU(2)$ subgroup of $E_8$ (a configuration that is possible by standard
existence theorems), breaking $E_8$ to $E_7$ and giving a low
energy spectrum that consists of six ${\bf 56}$'s of $E_7$.
(Note that as the ${\bf 56}$ is pseudoreal, it is possible for $E_7$
to act on 28 hypermultiplets transforming in the ${\bf 56}$ of $E_7$.
The spectrum consists of six copies of this.)  
Sequential Higgsing, turning on the expectation values of 
successive ${\bf 56}$'s,
can then be seen to completely break $E_7$.  For $k=9$, a similar
construction gives five ${\bf 56}$'s of $E_7$, and sequential
Higgsing now leads to a vacuum with an unbroken level one $SU(3)$.
Sequential Higgsing does not always give all the possible vacua,
as shown in an explicit example in \cite{Seiberg2}, and there may
be other branches, but  at any rate there is one branch of the $k=9$
moduli space in which $E_8$ is generically broken to $SU(3)$.} 
We note that complete Higgsing may be possible for $k=9$ in conformal
field theory even if it does not occur in classical geometry.
\footnote{Six-dimensional supersymmetry permits ``phase transitions''
in which branches of the moduli space of vacua with different generic
unbroken gauge groups meet at a point of enhanced gauge symmetry.
  For example, according to \cite{Seiberg2}, 
 a branch with generic unbroken $SU(3)$ and  a branch with generic
complete Higgsing can meet at a point at which the unbroken gauge symmetry
is $SU(6)$, with a hypermultiplet spectrum consisting of six ${\bf 6}$'s
of $SU(6).$  The necessary enhanced gauge symmetry might well occur
in conformal field theory but not in classical geometry.}

The generalization of (\ref{dimform}) for an arbitrary simple Lie group
$G$ with dual Coxeter number $h$ and dimension $
\dim\,G$ is
\be
\dim {\cal M}_k(G) = 4hk -4\,\dim\,G,
\la{genform}
\ee
valid whenever complete Higgsing is possible and in particular whenever
$k$ is sufficiently big.

For our problem of $E_8$ instantons with instanton number 12,
complete Higgsing is possible, and the gauge group is generically
completely broken.  On suitable loci in moduli space, with the property
that the instantons fit into a subgroup $H$ of $E_8$, a subgroup
of $E_8$ is restored -- namely the subgroup $G$ that commutes with $H$,
known as the commutant of $H$. 
When the vacuum gauge field reaches such a locus, perturbative
$G$ gauge fields will appear.  According to our discussion in the last
section, the un-Higgsing or restoration
of $G$ will be dual to the appearance, on
some other locus, of a non-perturbative gauge invariance with a gauge
group isomorphic to $G$.

Non-perturbative gauge fields, that is gauge fields that are not
seen in conformal field theory but appear no matter how weak the string
coupling constant may be, can only arise when the $K3$ or the vacuum gauge
bundle develops a singularity, causing string perturbation theory not to
be uniformly valid for all states.   We would like to find plausible
examples of how this works in practice.  That is, for suitable
groups $G$, we would like to identify the $K3$ or gauge singularity that
generates non-perturbative gauge invariance with gauge group $G$.
We will not be able to do this for all $G$, but we will find what
seem like compelling candidates for some simple cases.  Our discussion
is necessarily  incomplete, and no substitute for actually
understanding the map of hypermultiplet moduli space that appeared
in the last section.

It seems natural to first consider singularities of the gauge bundle
only, keeping the $K3$ smooth.  On a smooth $K3$, a singularity of the
gauge bundle (keeping the $E_8$ completely broken) necessarily consists
of a certain number of instantons shrinking to points.  For example,
consider the basic case in which a single instanton shrinks to a point.
The effective $k$ of the remaining gauge bundle diminishes by 1, so
according to (\ref{dimform}) the dimension of the $E_8$ instanton moduli space
drops by 120.  However, one is left with four parameters for the position
of the point instanton, so actually only $120-4= 116$ parameters
must be adjusted to make a single instanton shrink.

For the $SO(32)$ heterotic string, a single point instanton gives
\cite{Wittensmall} a non-perturbative $SU(2)$ gauge symmetry.
There is no general derivation of this for $E_8\times E_8$,
and it seems doubtful that it is true in general (as illustrated,
among other things, by the special role that $k=12$ is
about to play).  But fortune sometimes favors the brave, and let us ask
whether in our particular case, the collapse of an instanton might
be dual to an unbroken perturbative $SU(2)$ gauge symmetry.

\def\M{{\cal M}}
$E_8$ has a maximal subgroup $SU(2)\times E_7$.  The $SU(2)$ appearing
in such an $SU(2)\times E_7$ is a minimal or (in conformal field theory
language) ``level one'' embedding of $SU(2)$ in $E_8$, and its commutant
is $E_7$.   To get
an unbroken level one perturbative $SU(2)$ from one of the $E_8$'s, the vacuum
gauge bundle must fit into an $E_7$ subgroup.  As $E_7$ has $h=18$
and dimension 133, (\ref{genform}) gives $\dim\,{\cal M}_k(E_7)=72k-532$.
In particular, $\dim\,\M_{12}(E_7)=332$.  Using also 
(\ref{ugform}) and  the fact that $448-332=116$, one must adjust
116 parameters to get a perturbative unbroken $SU(2)$ gauge  symmetry.
The fact that this is the number of parameters needed to get a point
instanton strongly suggests that  at $k=12$ the shrinking
of an instanton to a point really is dual to 
the un-Higgsing of an $SU(2)$.

Fortified by this result, let us consider the possible occurrence of
{\it two} point instantons, first considering the generic case that
they are placed at distinct points on $K3$.  The number of parameters
that must be adjusted to get two point instantons is $2\cdot 116=232$.
If one point instanton gives an $SU(2)$ gauge symmetry, then
two disjoint point instantons should very plausibly give $SU(2)\times SU(2)$.
The dual should involve perturbative un-Higgsing of $SU(2)\times SU(2)$.
A level one embedding of $SU(2)\times SU(2)$ in $E_8$ has
commutant $SO(12)$.\footnote{Recall that $E_8$ has a maximal subgroup
$SO(16)$, which contains $SO(12)\times SO(4) = SO(12)\times SU(2)\times
SU(2)$.  Actually, $SO(16)$ admits another inequivalent level one
embedding of $SU(2)\times SU(2)$ (the commutant being $SO(8)\times SU(2)
\times SU(2)$), but the two embeddings are conjugate in $E_8$.}
As $SO(12)$ has $h=10$ and dimension 66, we get ${\rm dim}\,M_{12}(SO(12))
= 216$.  Since $448-216=232$, the expected 232 parameters are  needed
to restore an $SU(2)\times SU(2)$ subgroup of $E_8$.

Let us now consider the case of two point instantons at the {\it same}
point.  Since one must adjust four more parameters to make the positions
of the two point instantons in $K3$ coincide, the number of parameters
required is now $2\cdot 116+4=236$.  Two coincident point instantons must
give a gauge group that contains $SU(2)\times SU(2)$.  For the $SO(32)$
heterotic string, two coincident point instantons give gauge group
$Sp(2)=SO(5)$; let us make the ansatz that that is true here also.
A level one embedding of $SO(5)$ in $E_8$ has commutant $SO(11)$
(with $SO(5)\times SO(11)\subset SO(16)\subset E_8$).  As $SO(11)$
has $h=9$ and dimension 55, one gets ${\rm dim}\,\M_{12}(SO(11))=212$.
With $448-212=236$, the expected 236 parameters must be adjusted to see
an unbroken $SO(5)$ subgroup of $E_8$.

Now we move on to consider the case of {\it three} small instantons.
The number of parameters that must be adjusted to create three small
instantons at distinct points is $3\cdot 116=348$.  
(Here we are on shakier grounds, as we will assume that there is a branch
of the $k=9$ moduli space with complete Higgsing, and as explained
above we do not know this to be true.  This uncertainty will affect
many of the observations below.) 
The generic non-perturbative gauge group for three small instantons
should be $SU(2)^3$.  A level one embedding of $SU(2)^3$ in $E_8$
has commutant $SO(8)\times SU(2)$ (with $(SO(8)\times SU(2))\times SU(2)^3
=SO(8)\times SO(4)^2\subset SO(16)\subset E_8$).  Here there is the
new feature that $SO(8)\times SU(2)$ is not simple; we can place
$k_1$ instantons in $SO(8)$ and $k_2$ in $SU(2)$, with $k_1+k_2=12$.
Since $SO(8)$ has $h=6$ and dimension $ 28$ while $SU(2)$ has $h=2$
and dimension 3, the dimension of the $SO(8)\times SU(2)$ moduli space
is $24k_1+8k_2-124$.  We will make the ansatz of picking $k_1$ and $k_2$
to make this as large as possible subject to the constraint that
{\it bona fide} $SU(2)$ instantons on $K3$ of the given $k_2$ actually
exist.  That latter constraint forces $k_2\geq 4$,\footnote{For $SU(2)$,
(\ref{genform}) gives $\dim \,\M_k(SU(2))=8k-12$.  $k$ point instantons
would have a $4k$ dimensional moduli space.  Honest $SU(2)$ instantons
as opposed to collapsed point instantons exist only when $\dim\,\M_k(SU(2))$
exceeds the dimension of the moduli space of collapsed instantons,
that is when $8k-12>4k$ or $k>3$  The dimension-counting alone
does not give a rigorous argument here, but a rigorous argument
can be made using the index theorem for the Dirac operator in the instanton
field and a vanishing argument.} so to maximize the dimension of
the moduli space, we take $(k_1,k_2)=(8,4)$, and then we find
that $\dim\,\M_{(8,4)}(SO(8)\times SU(2))= 100$.  As $448-100=348$,
one must adjust the expected 348 parameters to restore a level one $SU(2)^3$
subgroup of $E_8$.  In future, when we meet instantons in a group
$H=H'\times SU(2)$, we will always set the $SU(2)$ instanton number to be
4, as in the calculation just performed.

One can similarly consider the case of three collapsed instantons that are not
at distinct points.  For instance, two collapsed instantons at one point and
one at a third point should give a non-perturbative gauge group $Sp(2)\times
SU(2)=SO(5)\times SU(2)$.  To obtain such a configuration, one must
adjust $3\cdot 116+4= 352$ parameters ($3\cdot 116$ to make three instantons
collapse and 4 to make two of the collapsed instantons appear at the same
point).  The commutant of a level one embedding of $SO(5)\times SU(2)$
in $E_8$ is $SO(7)\times SU(2)$ (via $SO(7)\times SU(2)\times SO(5)\times
SU(2) \subset SO(16)\subset E_8$).  Using the fact that $SO(7)$ has
$h=5$ and dimension 21, along with facts already exploited, we get
$\dim\,\M_{(8,4)}(SO(7)\times SU(2) ) = 96$.  With $448-96=352$, 
one must adjust the expected 352 parameters to restore a perturbative
$SO(5)\times SU(2)$ subgroup.

The last example of this kind is the case of three collapsed instantons
all at the same point in $K3$.  $3\cdot 116+2\cdot 4 = 356$ parameters
must be adjusted to achieve this situation.  Based on the result
of \cite{Wittensmall} for $SO(32)$, we may guess that the non-perturbative
gauge group for three coincident small instantons will be $Sp(3)$.
The commutant of a level one embedding of $Sp(3)$ in $E_8$ is 
$G_2\times SU(2)$.  (This was found by embedding $Sp(3)\subset SU(6)
\subset SO(12)
\subset SO(16)\subset E_8$ and then, by hand, determining that the commutant
of $Sp(3)$ was $G_2\times SU(2)$ -- a method that also gave the
decomposition used later 
of the $E_8$ Lie algebra under $Sp(3)\times G_2\times SU(2)$.)
With $G_2$ having $h=4$ and dimension 14, one finds that
$\dim\,\M_{(8,4)}(G_2\times SU(2)) = 92$; as $448-92=356,$ one must
adjust the expected 356 parameters to restore an $Sp(3)$ subgroup
of $E_8$.

For more than three small instantons, the residual $E_8$ instanton
would have instanton number $\leq 8$, so that the right hand side of
(\ref{dimform}) would be negative.  This actually means that the residual
instanton cannot completely break the $E_8$ symmetry, so that perturbative
as well as non-perturbative gauge fields will appear.  We have not
been successful in finding duals of configurations with perturbative
as well as non-perturbative gauge fields, and will not discuss this here.

Let us now look in somewhat more detail at the example with three
point instantons all at the same point in $K3$, giving a non-perturbative
$Sp(3)$ subgroup.  $Sp(3)$ has many subgroups that we have seen,
such as $Sp(2)\times SU(2)$, $SU(2)^3$, etc.  In the non-perturbative
description, one sees these subgroups of $Sp(3)$ by perturbing the
very exceptional configuration with three coincident point instantons
to a less exceptional -- but still singular -- configuration in which
the instantons are not all small or all coincident.
In the perturbative description, an $Sp(3)$ which has been restored
or un-Higgsed
can be broken down to a subgroup by turning back on the
 ordinary Higgs mechanism.

But when one thinks about doing so, a puzzle presents itself.  Apart from
subgroups of $Sp(3)$ that we have seen, there are other subgroups
of $Sp(3)$, such as $SU(3)$, $U(1)$, $U(1)\times SU(2)$, 
that do not appear anywhere on the
non-perturbative side (in any deformation of the configuration  
with the three coincident point instantons).  What prevents Higgsing
of $Sp(3)$ to $SU(3)$ or other unwanted groups?

To answer this question on the perturbative side, 
we need to know the $Sp(3)$ quantum numbers
of the massless hypermultiplets that appear when the $Sp(3)$
is un-Higgsed.  The adjoint representation of $E_8$ decomposes
under $G_2\times SU(2)\times Sp(3)$ as 
\be
({\bf 7},{\bf 1},{\bf 14})
\oplus ({\bf 1},{\bf 2},{\bf 14'})\oplus ({\bf 7},{\bf 2},{\bf 6}),
\la{agop}
\ee
plus the adjoint of $G_2\times SU(2)\times Sp(3)$;
here ${\bf 1}$ is a trivial representation, ${\bf 7}$, ${\bf 2}$, and
${\bf 6}$ are the defining representations of $G_2$, $SU(2)$, and $Sp(3)$,
of the indicated dimensions, and ${\bf 14}$ and ${\bf 14'}$ are the two
fourteen dimensional representations of $Sp(3)$ -- the ${\bf 14}$ is
the traceless second rank antisymmetric tensor, and the ${\bf 14'}$ is
a traceless third rank antisymmetric tensor.   The decomposition in 
(\ref{agop})
means that when  a level one perturbative $Sp(3)$       is unbroken, 
the massless hypermultiplets
will transform as a certain number of ${\bf 14}$'s, ${\bf 14'}$'s, and
${\bf 6}$'s, determined by index theorems.  No other representations of
$Sp(3)$ can appear, as they are absent in the adjoint representation of 
$E_8$.  For instance, the number of
${\bf 14}$'s will be the index of the Dirac operator on $K3$ with values in
the $({\bf 7},{\bf 1})$ of $G_2\times SU(2)$, and similarly for the
other representations.

Now, the ${\bf 14}$'s and ${\bf 6}$'s are the desired
representations that can Higgs $Sp(3)$ to the groups that one
actually sees on the non-perturbative side, and nothing else.  
For instance, with ${\bf 6}$'s 
alone,  one can Higgs only  
down to $Sp(2)$ or $Sp(1)=SU(2)$ -- two of the groups
that we actually encountered above.  Using also the ${\bf 14}$ one can
get the other desired subgroups of $Sp(3)$.  But the ${\bf 14}'$ would
make it possible to Higgs down to unwanted subgroups of $Sp(3)$ such
as $SU(3)$ or $U(1)$.

At this point we must recall that there are several loci on which
restoration of a  level one $Sp(3)$ occurs.  They are labeled by
the $G_2\times SU(2)$ instanton numbers $(k_1,k_2)$ with $k_1+k_2=12$.  
For general
$(k_1,k_2)$, the unwanted representation will appear.  But we found
the situation of three coincident point instantons to be dual to un-Higgsed
$Sp(3)$ with $(k_1,k_2)=(8,4)$.  When and only  when the instanton number is 4,
the Dirac index with values in the  
$(1,2)$ of $G_2\times SU(2)$ vanishes, 
\footnote{A short-cut to deduce this directly from (\ref{genform}) without
having to go back to the general index theorem is as follows.
{}From (\ref{genform}), $\dim\,{\cal M}_4(SU(3))=20$  but
also $\dim\,{\cal M}_4(SU(2)\times U(1))=20$ (in the latter case
one understands that the instantons are all in the $SU(2)$, and that
$h=0$ for $U(1)$).  As these numbers are equal, for instanton number four
an $SU(3)$ instanton on $K3$ automatically takes values in an $SU(2)\times 
U(1)$
subgroup, so that there are no zero modes in the ${\bf 2}$ of $SU(2)$ by
which one could deform to an irreducible $SU(3)$ instanton.}
and the dreaded ${\bf 14'}$ does
not arise.  Computing the other indices, one finds that 
the hypermultiplet spectrum consists of one ${\bf 14}$ and thirty-two
${\bf 6}$'s.  (For unclear reasons, but surely not coincidentally,
 this is the spectrum that appears in 
the ADHM construction of instantons with instanton number three
with gauge group $SO(32)$ on $\R^4$.) 
With this spectrum, the perturbative $Sp(3)$ can be Higgsed down precisely
to those subgroups that we found on the non-perturbative side.

\bigskip
\noindent{\it Inclusion Of K3 Singularities}

To go farther, since we have exhausted the list of 
singularities of a completely Higgsed gauge bundle on a generic $K3$,
a natural step is to look for non-perturbative gauge fields  whose
origin involves a $K3$ singularity.  This is a potentially vast subject 
(unless 
one
has a systematic point of view), with many possibilities to consider.
We will point out a few candidates.

Since we found the dual to an un-Higgsed $Sp(3)$, let us look for the
dual to a relatively small group containing $Sp(3)$, namely $SU(6)$.
The commutant of a level one $SU(6)$ is $SU(3)\times SU(2)$
($E_8$ has a maximal subgroup $SU(6)\times SU(3)\times SU(2)$).
The $SU(3)\times SU(2)$ instanton moduli space with instanton numbers
$(8,4)$ has dimension 84 (using the fact that $SU(3)$ has $h=3$ and
dimension eight); as $448-84=364$, 364 parameters must be adjusted to
restore a perturbative $SU(6)$ gauge symmetry.  We should try to interpret
this number in terms of a singular $K3$, since we have exhausted what
can be done with non-singular $K3$'s.  The simplest $K3$ singularity
is an $A_1$  singularity; in classical geometry one must
adjust three parameters to make such a singularity,
but to make the  conformal field theory singular, one must adjust
also a theta angle \cite{Aspinwall}, making four. 
Since one must adjust 120 parameters to make an instanton collapse
and fix its position, we are tempted to write $364=3\cdot 120+4$
and to propose that the dual of a perturbative $SU(6)$ is a $K3$ with an
$A_1$  singularity at which there are also three point instantons.

To check out this idea further, we note that in general, to get the $A_1$
 singularity with $k$  point instantons sitting on top of it,
one should expect to have to adjust
\be
w_k=120k+4
\la{funform}
\ee
parameters.  We have already considered the $k=3$ case.  Let
us look at $k=2$.  Here we have two coincident point instantons at the 
singularity.  The gauge group is at least $Sp(2)$, which we would
get with two coincident instantons without the $A_1$ singularity.
A relatively small group containing $Sp(2)$ is $SU(4)$; let us ask
if this is the right group for two point instantons on top of an $A_1$
singularity.  In fact, the commutant of $SU(4)$ is $SO(10)$
(think of the chain $SU(4)\times SO(10)= SO(6)\times SO(10)\subset
SO(16)\subset E_8$).  $SO(10)$ has $h=8$ and dimension 45,
and $\dim\, \M_{12}(SO(10))= 204$.  Thus $ 448-204=244$ parameters
must be adjusted to restore a perturbative $SU(4)$ symmetry, in agreement
with the $k=2$ case of (\ref{funform}).

One might ask about the $k=1$ case of (\ref{funform}).  We were not able
to find a group  whose restoration involves  adjusting 124 parameters.
However, note that the results so far can be expressed by stating that
an $A_1$ singularity that captures $k$ point instantons gives an
$SU(2k)$ gauge group.  If we suppose that that holds also for $k=1$, then
the $A_1$ singularity with one point instanton gives an $SU(2)$ gauge group,
which is the same gauge group we obtained for one point instanton
without the $A_1$ singularity.  So we postulate that with only
one point instanton, the coincidence with an $A_1$ singularity leads
to no enhancement of the gauge group.  In essence, for each $k$, the
$k$ coincident point instantons give $SU(2k)$ or $Sp(k)$ depending
on whether or not they lie at an $A_1$ singularity, 
but the fact that  $SU(2)=Sp(1)$ means that
for $k=1$ the $A_1$ singularity gives no enhanced gauge symmetry.

If we  feel bold, we can observe that since in classical geometry
an $A_1$ singularity is the same as a $\Z_2$ orbifold singularity,
it is possible for an $A_1$ singularity to capture a {\it half-integral}
number of point instantons.  Why not then try to use (\ref{funform}) for
half-integral $k$?  This turns out to work perfectly, though we find this
somewhat puzzling for a reason stated below.  Since $k$ point instantons
on the $A_1$ singularity gave an $SU(2k)$ gauge group at least for
$k=2,3$, we compare the $k=5/2$ and  $ k=3/2$ cases to $SU(5)$ and $SU(3)$,
respectively.  The commutant of $SU(5)$ is $SU(5)$
($E_8$ has a maximal subgroup $SU(5)\times SU(5)$).  
Since $SU(5)$ has  $h=5$ and dimension 24,
$\dim \M_{12}(SU(5)) = 144$; as $448-144=304$, 
304 parameters must be adjusted to restore a perturbative $SU(5)$, in 
agreement with the $k=5/2$ case of (\ref{funform}).  Likewise, $SU(3)$ has
commutant $E_6$, which has $h=12$ and dimension 78, so $\dim \,\M_{12}
(E_6)=264$.  With $448-264=184$, one must adjust 184 parameters to
restore an $SU(3)$ gauge symmetry, in agreement with the $k=3/2$ case
of (\ref{funform}).  Now, however, we must confess to what is unsettling
about these ``successes'' for $SU(3)$ and $SU(5)$: in classical
geometry (\ref{funform}) is not valid for half-integral $k$ (there is
a correction involving an eta-invariant), and we do not know why
this simple formula seems to work in string theory.

\def\bar{\overline}
We have encountered many but not all subgroups of $SU(6)$ in this discussion.
Just as in the discussion of $Sp(3)$, we should ask on the perturbative
side to what  subgroups the $SU(6)$ can be Higgsed.
For this we need the fact that under $SU(3)\times SU(2)\times SU(6)$
the adjoint of $E_8$ decomposes as $({\bf 3},{\bf 2},{\bf \bar 6})\oplus
({\bf \bar 3},{\bf 2}, {\bf 6}) \oplus 
({\bf 3},{\bf 1},{\bf 15})\oplus ({\bf \bar 3},{\bf 1},{\bf \bar{15}})
\oplus ({\bf 1},{\bf 2},{\bf 20})$ (plus the adjoint).  The $SU(6)$ 
content will consist of a certain number of ${\bf 20}$'s as well
as ${\bf 6}$'s and ${\bf 15}$'s and their complex conjugates.  
The ${\bf 20}$'s are unwanted as they would enable Higgsing to subgroups
of $SU(6)$ that would not have an interpretation in terms of a perturbation
of an $A_1$ singularity with three point instantons.  Happily,
because the instanton number in the $SU(2)$ is four, the Dirac index
with values in the $({\bf 1},{\bf 2})$ of $SU(3)\times SU(2)$ vanishes, the
same lucky fact we used in the $Sp(3)$ discussion,
so again the unwanted  representation does not occur.  

The actual spectrum thus
consists only of ${\bf 6}$'s, ${\bf 15}$'s, and their conjugates.
With these representations, one can break $SU(6)$  only to groups
that have natural interpretations using the above ideas. ($SU(5)$ and
$SU(3)$ do occur -- they can be reached by Higgsing with ${\bf 6}$'s --
so the $k=5/2$ and $k=3/2$ cases of (\ref{funform}) are needed.)  For
instance, with the ${\bf 15}$ one can Higgs down to $SU(4)\times SU(2)$,
which corresponds to two point instantons at an $A_1$ singularity
and one somewhere else.  The commutant of $SU(4)\times SU(2)$ is
again $SU(4)\times SU(2)$.  As $\dim \,\M_{(8,4)}(SU(4)\times SU(2))
=88$, one must adjust $448-88=360$ parameters to see a perturbative
$SU(4)\times SU(2)$ gauge symmetry.  We write $360=244+116$ and
propose that in the dual interpretation, one adjusts 244 parameters
to get two point instantons at an $A_1$ singularity and 116 to get
one point instanton somewhere else.  Similarly, by using also a
${\bf 6}$, one can Higgs $SU(6)$ to $SU(3)\times SU(2)$, whose
commutant in $E_8$ is another $SU(6)$.  As $\dim\,\M_{12}(SU(6))
= 148$, one must adjust $448-148=300$ parameters to see a perturbative
$SU(3)\times SU(2)$.  On the non-perturbative side, we write 
$300=184+116$ and interpret this as 184 parameters to get an $A_1$ 
singularity that absorbs $3/2$ point instantons, and 116 to get
a point instanton somewhere else.

One more example of a similar kind, though not obtained by Higgsing
 of this particular $SU(6)$,
is to look for an unbroken level one $SU(3)\times SU(3)$ subgroup of $E_8$.
The commutant of $SU(3)\times SU(3)$ is another $SU(3)\times SU(3)$ 
($E_8$ has a subgroup $SU(3)^4$).  As $\dim \,\M_{k_1,k_2}(SU(3)\times
SU(3)) = 80$ (for any $k_1,k_2$ with $k_1+k_2=12$), one must adjust
$448-80=368$ parameters to get an unbroken perturbative $SU(3)\times
SU(3)$.  Writing $368=164+164$, we propose that the dual of this consists
of two disjoint $A_1$ singularities each of which has captured
$3/2$ point instantons.

We should note that similar examples with codimension very close to 448,
like the $k=7/2$ case of (\ref{funform}), do not seem to work.  We suspect
that this is because in these cases complete Higgsing to the expected
group is not possible.

\section{Higher Loops}

In this section, we will compare the loop expansion of the fundamental
string to that of the dual string.  The loop expansion
of the fundamental string 
for any given physical observable is an expansion of that observable
in 
powers of $e^\Phi$, valid as as aymptotic expansion near $\Phi=-\infty$.
The expansion takes the general form $\sum_{n\geq n_{0}}b_n \,e^{n\Phi}$
where $n_0$ depends on what observable is considered.  
(In the Einstein metric, the exponents may not be integers.)
The perturbation
expansion of the dual string is an analogous asymptotic expansion
in powers of $e^{-\Phi}$, valid near $\Phi=+\infty$.  The general form
is $\sum_{m\geq m_{0}}c_me^{-m\Phi}$.  We would like to make a term-wise
comparison of these expansions, but in general such a term-wise comparison
of asymptotic expansions of a function about two different points is not valid.
One situation in which such a termwise comparison {\it is} valid is
the case that each series has only finitely many terms; thus
given an equality $\sum_{n=n_0}^{n_1}b_ne^{n\Phi}=
\sum_{m=m_0}^{m_1}c_me^{-m\Phi}$ between finite sums,
 the exponents and coefficients must be equal.

An important reason for such a series to have only finitely many terms
is that supersymmetry may allow only finitely many terms in the expansion
of a given observable in powers of $e^{\pm \Phi}$.  For instance, in this
paper we have exploited the fact that low energy 
 gauge field kinetic energy
has a $\Phi$-dependence with only two possible terms.  In what follows,
we will work out the consequences of a term-wise comparison of the 
two perturbation expansions in any situation in which such a comparison
is valid, including
but not limited to the case in which supersymmetry allows only finitely
many terms.  

The fundamental string involves two kinds of loop expansion:  quantum $D =
6$ string loops ($L$) with loop expansion parameter $\alpha'e^{\Phi}$ and
classical $2$-dimensional $\sigma$-model loops with loop expansion parameter
$\alpha'$. Following \cite{Luloop}, let us consider the purely gravitational
contribution to the resulting effective action, using the string $\sigma$-model
metric: 
\be
{{\cal L}_{L,L+m} = a_{L,L+m} \frac{(2\pi)^3}{\alpha'^2} \sqrt{-G} e^{-\Phi}
\Bigg({\alpha' e^{\Phi}}\Bigg)^L
\alpha'{}^m R^n ,}
\la{graveffecact}
\ee
where $R^n$ is symbolic for a scalar contribution of $n$ Riemann tensors each
of dimension two.  One could also include covariant derivatives of $R$
(or other interactions with a known transformation law under duality), but
for our purposes (\ref{graveffecact}) will be sufficient.  The $a_{L,L+m}$
are numerical coefficients, not involving $\pi$. The subscripts $L$ and
$L+m$ have been chosen in anticipation of the relation $L+m=\tilde L$, with 
$\tilde L$ the number of loops in the dual theory.
Since  
\be
{[{\cal L}_{L,L+m}]=6,  \qquad
[\alpha'] = -2,}
\la{cannodim}
\ee
we have, on dimensional grounds,
\be
{m =  n - 1 - 2 L.}
\la{looprelation}
\ee

Likewise, the theory of the  dual string involves quantum $D=6$ dual string
loops ($\tilde L$) with loop expansion parameter ${\alpha'
e^{-\Phi}}$ and classical $2$-dimensional $\sigma$-model loops
with loop expansion parameter $\alpha'$. The corresponding Lagrangian
using the dual $\sigma$-model metric is  
\be
{\tilde{{\cal L}}_{\tilde{L}+\tilde{m}, \tilde{L}} = \tilde{a}_{
\tilde{L}+ \tilde {m}, L}
 \frac{(2\pi)^3}{\alpha'^2}\sqrt{-\tilde G} e^{\Phi}
\bigg({\alpha'e^{-\Phi}}
\bigg)^{\tilde{L}}  \alpha'{}^{\tilde{m}} \tilde R^n .}
\la{dugraeffec}
\ee
Again, on dimensional grounds,
\be
{\tilde{m} = n - 1 - 2 \tilde{L}.}
\la{duloprelat}
\ee
Our fundamental assumption is that ${\cal L}$ and $\tilde{\cal L}$ are
related by duality which implies, in particular, that the purely
gravitational contributions should be identical when written in the same
variables.  So from (\ref{stringmetric}) and (\ref{dualstringmetric}) and
transforming to the canonical metric, but dropping the c superscript, we
find:  
 
\be
{{\cal L}_{L,L+m} =  \frac{(2\pi)^3}{\alpha'^2}a_{L,L+m}
\alpha'^{L+m} e^{-m \Phi/2} \sqrt{-G} {R}^n,}
\la{canoact}
\ee

\be
{\tilde{{\cal L}}_{\tilde{L}+\tilde{m},\tilde{L}}}=\frac{(2\pi)^3}{\alpha'^2}
\tilde{a}_{\tilde{L}+\tilde{m},\tilde{L}}
\alpha'{}^{\tilde L + \tilde{m}} e^{\tilde{m} \Phi/2} \sqrt{-G} {R}^n 
\la{ducanoact} 
\ee  
where we have also dropped the dilaton derivative terms. We find
that ${\cal L}$ and $\tilde{{\cal L}}$ do coincide provided   
\be
{m + \tilde{m} = 0,}
\la{mtildem}
\ee
i.e from (\ref{looprelation}) and (\ref{duloprelat}), provided
\be
{m = \tilde{L} - L = - \tilde{m},}
\la{lmrelation}
\ee
\be
{n = L + \tilde{L} + 1,}
\la{nlrelation}
\ee
with
\be
{a_{L,\tilde{L}} = \tilde{a}_{L ,\tilde{L}}.}
\la{atildea}
\ee
 Hence the total Lagrangian can be elegantly written
\[
{{\cal L} = \sum\limits_{L,\tilde L} {\cal L}_{L,\tilde{L}}.}
\]
\be
{{\cal L}_{L,\tilde{L}} = a_{L,\tilde{L}}
{(2\pi)^3} \alpha'^{L+\tilde{L}-2}
e^{(L-\tilde{L}) \Phi/2} \sqrt{-G}
{R}^{L + \tilde{L} +1}.}
\la{resultact}
\ee
Thus we see that under heterotic string/string duality, the worldsheet loop
expansion of one string corresponds to the spacetime loop expansion of the
other. Moreover, (\ref{resultact}) gives rise to an infinite number of
non-renormalization theorems.  
(As explained at the beginning of this section, these theorems hold
if it is known {\it a priori} that each perturbation expansion has only
finitely many terms, and perhaps under wider but presently unknown
hypotheses.)
The first of these is the absence of a
cosmological term $\sqrt{-G} {R}^0$. The second states that $\sqrt{-G}R^1$
appears only at ($L = 0, \tilde{L} = 0$) and hence the tree level action does
not get renormalized. The third states that  $\sqrt{-G}R^2$ appears only at ($L
= 0, \tilde{L}  =1$) and ($L = 1, \tilde{L}  =0$), and so on.  Since $F$ has
the same dimension as $R$, similar restrictions will apply to the pure
Yang-Mills and mixed 
gravity-Yang Mills Lagrangians. The ($L = 0, \tilde{L} = 0$), ($L
= 0, \tilde{L}  =1$) and  ($L = 1, \tilde{L} = 0$) terms correspond 
respectively
to the $I_{00}$, $I_{01}$ and $I_{10}$ of the previous section. Self-duality,
given by  (\ref{discrete}), imposes the further constraint   \be  
a_{L ,\tilde L}=a_{\tilde L,  L}
\ee
and means that the spacetime and worldsheet loop expansions are in
fact identical.

In the situation in which the term-by-term comparison of the two expansions
is justified by a supersymmetry argument showing that each expansion has
only finitely many terms, it would be interesting to compare the
restrictions on exponents following from supersymmetry with those
that follow from duality.  In principle, duality might give a restriction
more severe or less severe than the one that follows from duality,
but in the few familiar cases the two restrictions agree.

\section{Acknowledgements}

MJD and RM have 
benefitted from conversations with Joachim Rahmfeld and Jim Liu.

\appendix

\section{The String Tension And The Dual String Tension}

We will here explore from an eleven-dimensional point of view what is
entailed in setting the fundamental and dual string tensions equal.
Since the string is obtained by wrapping a membrane of tension $T_3$ around
$S^1/Z_2$ with radius $R$ and the dual string is obtained by wrapping a
fivebrane of tension ${\tilde T}_6$ around $K3$ of volume $V$, the string
tension ${\hat T}$ and dual string tension ${\hat {\tilde T}}$, measured in
the $D=11$ metric are given by     
\[
{\hat T}=RT_3
\]
\be
{\hat {\tilde T}}=V{\tilde T}_6
\ee
Let us denote length scales measured in the $D=11$ metric $\hat G$, $D=6$ 
string
metric $G$ and $D=6$ dual string metric $\tilde G$ as $\hat L$, $L$ and $\tilde
L$, respectively. Since
\[
\hat G= R^{-1} G
\]
\be
\hat G=V^{-1} \tilde G
\ee
they are related by
\[
L^2=R{\hat L}^2
\]
\be
{\tilde L}^2=V{\hat L}^2
\ee
Since the string tension measured in the string metric, $T$, and the dual 
string
tension  measured in the dual string metric, $\tilde T$, both have dimensions
$(length)^{-2}$, they are given by 
\[
T=R^{-1}{\hat T}=T_3
\]
\be
\tilde T=V^{-1}{\hat {\tilde T}}={\tilde T}_6
\ee
and are therefore both independent of $R$ and $V$. In fact, since \cite{Liu} 
\be
{\tilde T}_6 \sim T_3{}^2
\ee
we may, without loss of generality, choose units for which $T$ and $\tilde T$
are equal.

\section{A Note On The $v$'s and $\tilde v$'s}

An important feature of the
heterotic/heterotic duality conjecture concerns the numerical coefficients
$v_\alpha$ appearing in $X_4$ and $\tilde v_\alpha$  appearing in $\tilde X_4$.
Here we wish to make a remark on the precise normalization of these
coefficients.
In string perturbation theory each $v$ is given by \cite{Schellekens} 
\be
v \tr F^2=\frac{n}{h}\Tr F^2
\ee
where $n$ is the level of the Kac-Moody algebra, $h$ is the dual Coxeter
number, $\tr$ denotes the trace in the fundamental representation and $\Tr$ 
is the trace in
the adjoint representation.  For example,  
\be
\begin{array}{lclclcl}
h_{SU(N)}&=&N&&\Tr F_{SU(N)}{}^2&=&2N\tr F_{SU(N)}{}^2\\
h_{SO(N)}&=&N-2&&\Tr F_{SO(N)}{}^2&=&(N-2)\tr F_{SO(N)}{}^2\\
h_{Sp(N)}&=&N+1&&\Tr F_{Sp(N)}{}^2&=&(2N+2)\tr F_{Sp(N)}{}^2\\
h_{G_2}&=&4&&\Tr F_{G_2}{}^2&=&4\tr F_{G_2}{}^2\\
h_{F_4}&=&9&&\Tr F_{F_4}{}^2&=&3\tr F_{F_4}{}^2\\
h_{E_6}&=&12&&\Tr F_{E_6}{}^2&=&4\tr F_{E_6}{}^2\\
h_{E_7}&=&18&&\Tr F_{E_7}{}^2&=&3\tr F_{E_7}{}^2\\
h_{E_8}&=&30&&\Tr F_{E_8}{}^2&=&30\tr F_{E_8}{}^2
\end{array}
\ee
The values of $h$ were used in section (4).

So, for $G=E_7,~E_6,~SO(10),~SU(5)$ one finds $v_{G}=n/6,~n/3,~n,~2n$,
respectively. In the case of $SO(32)$ with an $k=24$ embedding,
one finds from (\ref{SO(32)}) that $(n_{SO(28)}=1,n_{SU(2)}=1)$ but $(\tilde
n_{SO(28)}=\mp 2,\tilde n_{SU(2)}=\pm 22)$ and hence we encounter the
``wrong-sign problem''\footnote{These $\tilde n$ differ from the {\it
shifted} dual Kac-Moody levels defined in \cite{Minasian}.}.  A similar
problem arises for $E_8 \times E_8$ except in the case of symmetric 
embedding where both $k_i=12$. For generic embeddings one finds that $n_i=1$
but      
\be   
\tilde n_i=\frac{1}{2}(k_i-12)  
\ee
Since one is limited to  $k_1+k_2=24$, 
one factor will always have the wrong
sign except when for the $k=12$ case discussed in this paper, for which
the $\tilde n_i$ both vanish.

%%%\bibliographystyle{preprint} %%%\bibliography{duality}
%%%%%%%%%%%%% include bibtex generated bibliography %%%%%%%%%%%%%%%

%%%%%%%%%%%%%%%%%% end bibliography %%%%%%%%%%%%%%%%%%%%%%%%%%%%%%%

\end{document}